# About inevitability of budgetary code
# Receiving for fiscal politics

George Abuselidze

**Abstract**

Since the end of 90s till today when all the elements confirming Georgian State System have practically been established, budget system and policy remains as the most difficult Georgian macroeconomics challenge and even still half-and-half unsolved problem. One side of the fiscal policy is quite crucially formulated and administrative Tax Code, and the other side is the weak, unmanaged and incomplete law on "Budget System".

The formation of the optimal economic and social organization model of the country is essentially dependent on the correct fiscal policy implementation.

According to the above-mentioned the elaboration and adoption of the Budget Code having equal force as Tax Code is necessary by which the following are to be determined: excellence of government responsibility when it will not perform the budget obligations specified by the law permanently; the rights and responsibilities of the state, the optimal distribution of the funds mobilized by the tax towards each member of the society.

For optimization of the budget system effective correlation between the state, regional and local budgets revenues and expenditures is particularly important as the social-economic development of the regions and territorial units of the country is impossible without the financial relations. For it the just differentiation of tax base in the section of state, regional and local budgets and transfers system for support of the budgets of the territorial units from the central budget are necessary.

Solving the most part of these problems is possible by the adoption of the budget code which, in our opinion, is to be considered as the closest decisive task for the current legislative and executive authority. Therefore the adoption of the budget code should not be end in itself, today the most significant challenge before Georgian state is to apply the budget code so that not to create the ground of destabilization of the financial system of the country and not to lose its regulating role in the formation and development of the united economical and financial system of the country.

# საგადასახადო კოდექსის ადეკვატური საბიუჯეტო კოდექსის მიღების აუცილებლობის შესახებ

გიორგი აბუსელიძე,
ბათუმის დამოუკიდებელი უნივერსიტეტის პროფესორი

**აბსტრაქტი**

90-იანი წლების ბოლოდან დღემდე, როდესაც საქართველოს სახელმწიფოებრიობის დამადასტურებელი ყველა ელემენტი პრაქტიკულად ჩამოყალიბდა, ბიუჯეტური სისტემა და საბიუჯეტო პოლიტიკა ქართული მაკროეკონომიკის ურთულეს და ჯერ კიდევ სანახევროდაც გადაუჭრელ პრობლემად რჩება. ფისკალური პოლიტიკის ერთ მხარეს არის საკმაოდ მკაცრად ფორმულირებული და ადმინისტრირებადი საგადასახადო კოდექსი, ხოლო მეორე მხარეს – სუსტი, გაუმართავი და არასრულყოფილი კანონი "საბიუჯეტო სისტემის შესახებ".

ქვეყნის ოპტიმალური ეკონომიკური და სოციალური მოწყობის მოდელის ფორმირება არსებითად დამოკიდებულია სწორი ფისკალური პოლიტიკის განხორციელებაზე.

აღნიშნულიდან გამომდინარე, საჭიროა საგადასახადო კოდექსის საპირწონე საბიუჯეტო კოდექსის შემუშავება და მიღება, რომლითაც უნდა განისაზღვროს: ხელისუფლების პასუხისმგებლობის ხარისხი როდესაც იგი პერმანენტულად არ შეასრულებს კანონით განსაზღვრულ საბიუჯეტო ვალდებულებებს; სახელმწიფოს უფლება-მოვალეობები, გადასახადით მობილიზებული ფულადი სახსრების საზოგადოების თითოეული წევრის მიმართ ოპტიმალური გადანაწილების მიმართულებით.

ქვეყნის ბიუჯეტური სისტემის ოპტიმიზაციისათვის განსაკუთრებით მნიშვნელოვანია სახელმწიფო, რეგიონული და ადგილობრივი ბიუჯეტების შემოსავლებსა და გასავლებს შორის ეფექტიანი თანაფარდობა, რამეთუ ქვეყნის რეგიონების, ტერიტორიული ერთეულების სოციალურ-ეკონომიკური განვითარება შეუძლებელია საფინანსო ურთიერთობათა გარეშე. ამისათვის კი აუცილებელია: საგადასახადო ბაზის სამართლიანი დიფერენციაცია სახელმწიფო, რეგიონული და ადგილობრივი ბიუჯეტების ჭრილში; ტრანსფერტების სისტემა ცენტრალური ბიუჯეტიდან ტერიტორიული ერთეულების ბიუჯეტების მხარდასაჭერად.

ამ პრობლემათა უდიდესი ნაწილის გადაჭრა შესაძლებელია საბიუჯეტო კოდექსის მიღებით, რომელიც, ჩვენი აზრით, დღევანდელი საკანონმდებლო და აღმასრულებელი ხელისუფლების უახლოეს გადასაწყვეტ ამოცანად უნდა იქნეს მიჩნეული. ამასთანავე საბიუჯეტო კოდექსის მიღება თვითმიზანი არ უნდა იყოს, საქართველოს სახელმწიფოს წინაშე დღეისათვის უმთავრესი ამოცანაა გამოყენებულ იქნას საბიუჯეტო პოლიტიკა ისეთნაირად, რომ არ მიეცეს საფუძველი ქვეყნის ფინანსური სისტემის დესტაბილიზაციას, არ დაიკარგოს მისი მარეგულირებელი როლი ქვეყნის ერთიანი ეკონომიკური და ფინანსური სისტემის ფორმირების და განვითარების საქმეში.

ძირითადი ნაწილი

თანამედროვე ეტაპზე საქართველოში მიმდინარე ეკონომიკური და პოლიტიკური სისტემის რადიკალური ტრანსფორმაციის პროცესი, ახალი ეკონომიკური სისტემის შექმნა და მისი ეფექტიანობის უზრუნველყოფა მოითხოვს კოლოსალური ფინანსური რესურსების მობილიზაციას, რაც შეუძლებელია ფინანსური მექანიზმის გამართული ფუნქციონირების გარეშე. ეს კი, თავის მხრივ საჭიროებს სათანადო საბიუჯეტო პოლიტიკის განხორციელებას.

სამწუხაროდ, საქართველოს ბიუჯეტური სისტემა და საბიუჯეტო პოლიტიკა მაკროეკონომიკის ურთულეს და ჯერ კიდევ სანახევროდაც გადაუჭრელ პრობლემად რჩება. ქვეყნისათვის თვისებრივად ახალი საფინანსო-საბიუჯეტო პოლიტიკის ფორმირებას საფუძველი დაუდო პირველი ეროვნული საბიუჯეტო საკანონმდებლო აქტის "საბიუჯეტო სისტემისა და უფლებების შესახებ" მიღებამ (30.03.1993), რომელიც 1996 წლის მაისში შეცვალა კანონმა "საბიუჯეტო სისტემისა და საბიუჯეტო უფლებამოსილებათა შესახებ". ეს თავის მხრივ შეცვლილ იქნა 2003 წლის 23 აპრილს მიღებული კანონით "საბიუჯეტო სისტემის შესახებ", რომელიც მინიმალურადაც არ აკმაყოფილებს ქვეყნის საბიუჯეტო პოლიტიკის წინაშე მდგარ თანამედროვე გაზრდილ მოთხოვნებს, იმ პოლიტიკური, ეკონომიკური და სოციალური ამოცანების დაძლევას, რომელიც დღის წესრიგში დააყენა საზოგადოებაში მიმდინარე რადიკალურმა ტრანსფორმაციულმა მოვლენებმა.

ქვეყნის ოპტიმალური ეკონომიკური და სოციალური მოწყობის მოდელის ფორმირება არსებითად დამოკიდებულია სწორი ფისკალური პოლიტიკის განხორციელებაზე. განვლილი წლების საბიუჯეტო საქმიანობის პრაქტიკამ (იხ. ცხრილი) გვიჩვენა, რომ ქვეყნის ბიუჯეტური სისტემა ვერ ასახავს გარდამავალი პერიოდის სოციალურ-ეკონომიკურ თავისებურებებს და მოითხოვს შემდგომ რადიკალურ ტრანსფორმაცია-სრულყოფას. მოქმედ ბიუჯეტურ სისტემაში მრავალი საბიუჯეტო სამართლებრივი ნორმის უკმარისობაა, რომელიც განსაზღვრავდა სახელმწიფო და ადგილობრივ ბიუჯეტებს შორის ურთიერთობას, ბიუჯეტური სისტემის რგოლებს შორის შემოსავლების განაწილებას, ბიუჯეტის შედგენის პრინციპებს, ხარჯების დაფინანსებას და სხვ. ყურადსადებია ასევე ტრანსფერტების განაწილებაც, რაც სახელმწიფოს მხრიდან არ ემყარება არავითარ სტრატეგიას და პოლიტიკას, ტრანსფერტების განაწილების წესი არ არის განსაზღვრული არც კანონით და არც რაიმე კანონქვემდებარე ნორმატიული აქტით. ობიექტური რეალობიდან გამომდინარე, დღეს საქართველოს სჭირდება სიცოცხლისუნარიანი ბიუჯეტური სისტემა, რომელმაც უნდა უზრუნველყოს ქვეყნის ეკონომიკური აღმავლობა.

საქართველოს ერთიანი საფინანსო-საბიუჯეტო პოლიტიკა, როგორც ქვეყნის ეკონომიკური პოლიტიკის შემადგენელი ნაწილი, ხორციელდება საგადასახადო, ფულად-საკრედიტო, ფასებისა და სავალუტო პოლიტიკასთან შეთანხმებულად. ისინი ერთმანეთთან არა მხოლოდ მჭიდროდ არის დაკავშირებული, ურთიერთს განაპირობებენ და ერთიან დიალექტიკურ მთლიანობას ქმნიან.

ქვეყნის საგადასახადო და საბიუჯეტო პოლიტიკის ოპტიმალურობაზე დიდადაა დამოკიდებული ეკონომიკური ზრდის ტემპები. საქართველოში უკვე საკმაოდ დიდი ხანია მოქმედებს საგადასახადო კოდექსი, რომელმაც ასე თუ ისე ჩამოაყალიბა სამეურნეო სუბიექტების ვალდებულება და პასუხისმგებლობა სახელმწიფოს წინაშე.

საგადასახადო კოდექსის თანახმად, სახელმწიფო თავის ფონდში (ბიუჯეტში) ითხოვს შემოსავლების ნაწილის კანონიერად ჩარიცხვას, მეორე მხრივ კი, სახელმწიფო იღებს ვალდებულებას (ბიუჯეტის შესახებ მიღებული ყოველწლიური კანონით) მობილიზებული შემოსავლები გადაანაწილოს სახელმწიფოს და მოსახლეობას შორის პოლიტიკური, ეკონომიკური და სოციალური ფუნქციების შესაბამისად, თუმცა ამ ფუნქციათა შეუსრულებლობის შემთხვევაში იგი პასუხისმგებლობას არ კისრულობს. მაგრამ იმ შემთხვევაში, თუ გადასახადის გადამხდელმა თავი აარიდა გადახდას და დამალა შემოსავლები, სახელმწიფო ამუშავებს მის მიერ მიღებულ კანონმდებლობას (საგადასახადო კოდექსს და სხვა კანონებს), რაც ითვალისწინებს შემდეგ დირექტივა-განაჩენს: "უნდა გადაიხადო! თუ არ გადაიხდი, ვერ იარსებებ!". ამრიგად, სახელმწიფო როგორც ბუნებრივი მონოპოლისტი, გადასახადების ოდენობის დადგენასთან ერთად, საჭიროების შემთხვევაში, მის ამოსაღებად იყენებს იძულებით მეთოდებს, ხოლო, მეორე მხრივ, მის მიერვე დადგენილი ვალდებულებების (ხარჯების გეგმის) შეუსრულებლობის შემთხვევაში, არანაირ პასუხისმგებლობას არ იღებს.

ამრიგად, ფისკალური პოლიტიკის ერთ მხარეს არის საკმაოდ მკაცრად ფორმულირებული და ადმინისტრირებადი საგადასახადო კოდექსი, ხოლო მეორე მხარეს – სუსტი, გაუმართავი და არასრულყოფილი კანონი "საბიუჯეტო სისტემის შესახებ".[1] ამ "ეკონომიკური სასწორის" ორივე მხარეს იმყოფება სახელმწიფო, მისი საკანონმდებლო და აღმასრულებელი სტრუქტურებით.

ყოველივე ზემოთაღნიშნული მეტყველებს არათანაბარ, უსამართლო გარემოზე სახელმწიფოსა და გადასახადის გადამხდელისათვის, პირველის სასარგებლოდ, რაც ხშირად წარმოშობს ფარულ (ზოგჯერ კი ღია) დაპირისპირებას მხარეებს შორის.

ბიუჯეტი, საბიუჯეტო სისტემა და მთლიანად საბიუჯეტო პროცესი არის ის უმთავრესი რგოლი, რომელმაც უნდა დაამყაროს წონასწორობა ხელისუფლებასა და საზოგადოებას შორის. და თუ ჩვენი მიზანია მძლავრი ეროვნული ეკონომიკის შექმნა, ხელისუფლებასა და მეწარმეებს შორის გაუთავებელი დაპირისპირების დასრულება, საქართველოს გამოყვანა უმწვავესი საფინანსო-ეკონომიკური და სოციალურ-პოლიტიკური კრიზისიდან, აუცილებელია საგადასახადო კოდექსის ადეკვატური საბიუჯეტო კოდექსის შემუშავება და მიღება, რომლითაც უნდა განისაზღვროს: ხელისუფლების პასუხისმგებლობის ხარისხი, როდესაც იგი პერმანენტულად არ შეასრულებს კანონით განსაზღვრულ საბიუჯეტო ვალდებულებებს; სახელმწიფოს უფლება-მოვალეობები გადასახადით მობილიზებული ფულადი სახსრების საზოგადოების თითოეული წევრის მიმართ ოპტიმალური გადანაწილების მიმართულებით.

ფისკალურ პოლიტიკაში ერთ-ერთ რთულ პრობლემას სხვადასხვა დონის ბიუჯეტებს შორის შემოსავლების განაწილება წარმოადგენს. კერძოდ, ადგილობრივ ბიუჯეტებში მეტწილად ჩაირიცხება ძნელად ამოსადები და მცირე ოდენობის საგადასახადო შემოსავლები, რაც მძიმე პირობებს უქმნის ადგილობრივ ორგანოებს. ეს გამოწვეულია იმით, რომ ჯერ კიდევ არ არის შემუშავებული საგადასახადო შემოსავლების სამართლიანი განაწილების მეთოდოლოგია და მეთოდიკა.

---

[1] გ. აბუსელიძე, ბიუჯეტური სისტემის ფორმირების სამართლებრივი საფუძვლების ოპტიმიზაციისათვის, ჟურნ. "სოციალური ეკონომიკა", №4, 2005 წ, გვ. 72.

ქვეყნის ბიუჯეტური სისტემის ოპტიმიზაციისათვის განსაკუთრებით მნიშვნელოვანია სახელმწიფო, რეგიონული და ადგილობრივი ბიუჯეტების შემოსავლებსა და ხარჯებს შორის ეფექტიანი თანაფარდობა, რამეთუ ქვეყნის რეგიონების, ტერიტორიული ერთეულების სოციალურ-ეკონომიკური განვითარება შეუძლებელია ფინანსურ ურთიერთობათა გარეშე. ამისათვის კი აუცილებელია საგადასახადო ბაზის სამართლიანი დიფერენციაცია სახელმწიფო, რეგიონული და ადგილობრივი ბიუჯეტების ჭრილით, რამეთუ მხოლოდ ქონების გადასახადი ვერ იქნება ადგილობრივი ბიუჯეტების შემოსავლების ფორმირების მნიშვნელოვანი წყარო. ამის გარეშე კი ადგილობრივი ბიუჯეტები ყოველთვის დამოკიდებული იქნება ზემდგომ საბიუჯეტო დონეებზე, რაც ეწინააღმდეგება ბიუჯეტის შედგენის ერთ-ერთ მნიშვნელოვან პრინციპს – "დამოუკიდებლობის პრინციპს". ამისათვის კი აუცილებელია ტრანსფერტების სისტემა ცენტრალური ბიუჯეტიდან - ტერიტორიული ერთეულების ბიუჯეტების მხარდასაჭერად, დამატებითი ტრანსფერტები დოტაციებისა და სუბსიდიების სახით, მიზნობრივი ფინანსური მხარდაჭერა (ენერგეტიკის სფერო, მთიანი რაიონები და ა.შ.) და სხვ.

ამრიგად, ეროვნული ეკონომიკა და მისი ფინანსური სფერო სტაბილური და უსაფრთხოა იმ შემთხვევაში, როდესაც იგი ძირითადად დამოკიდებულია საკუთარ ეკონომიკურ და ფინანსურ რესურსებზე, ხოლო ბიუჯეტის დეფიციტის დასაფარავად უცხოური წყაროები გამოყენებულია როგორც დამხლევი სისტემა, რაც გულისხმობს, რომ დაფინანსების საგარეო წყაროების ხვედრითი წონა ბიუჯეტის დეფიციტის დაფინანსებაში არ უნდა აღემატებოდეს მაქსიმუმ 15-20%-ს და მთლიანობაში ბიუჯეტის დეფიციტმა არ უნდა გადააჭარბოს მშპ-ს 2%-ს. ქვეყნის საფინანსო-ეკონომიკურ უსაფრთხოებისას, უპირველეს ყოვლისა, საყურადღებოა დამოკიდებულება მშპ-ის დინამიკასა და ბიუჯეტის ხარჯების სტრუქტურაში საგარეო ვალების მომსახურების წილს შორის. როდესაც მშპ-ის მოცულობა კლებულობს და შესაბამისად მცირეა საბიუჯეტო შემოსავლებიც, ბუნებრივია, ვალების მომსახურებაზე წარიმართება საბიუჯეტო შემოსავლების უფრო მეტი წილი, მაგრამ როდესაც მშპ-ის მოცულობა იზრდება და შესაბამისად მატულობს საბიუჯეტო მაჩვენებლების აბსოლუტური პარამეტრები, ცხადია, ფიქსირებული საგარეო ვალის მომსახურებისათვის საჭირო საბიუჯეტო ასიგნება ნაკლებ ადგილს იკავებს საბიუჯეტო შემოსავლებში.

საქართველოს სახელმწიფოს უმთავრესი სადღეისო ამოცანაა ისეთი საბიუჯეტო პოლიტიკის განხორციელება, რაც გამორიცხავს ქვეყნის ფინანსური სისტემის დესტაბილიზაციას, გააძლიერებს ადეკვატური პოლიტიკის მარეგულირებელ როლს ქვეყნის ერთიანი ეკონომიკური და ფინანსური სისტემის ფორმირებას და განვითარებაში.

ამასთანავე, საბიუჯეტო კოდექსის მიღება თვითმიზანი არ უნდა იყოს და მისი შემუშავებისას არ უნდა განმეორდეს საგადასახადო კოდექსის მსგავსი ხარვეზები და ნაკლოვანებები, რომლებიც მიუხედავად მრავალჯერადი და მრავალრიცხოვანი ფრაგმენტული ცვლილებისა (4000-მდე ცვლილებაა შეტანილი), დღემდე არ არის სრულყოფილი, რის გამოც სამართლიანად არის მიჩნეული არასრულყოფილ საკანონმდებლო დოკუმენტად.

გამომდინარე ყოველივე ზემოთაღნიშნულიდან, მიგვაჩნია, რომ ქვეყნის ფისკალურ პოლიტიკაში არის რიგი გადაუჭრელი პრობლემა, რაც საფრთხეს უქმნის საფინანსო-ეკონომიკურ სტაბილურობას. ამ პრობლემათა უდიდესი ნაწილის გადაჭრა შესაძლებელია გამართული, მეცნიერულად დასაბუთებული საბიუჯეტო კოდექსის მიღებით, რაც ქვეყნის საკანონმდებლო და აღმასრულებელი ხელისუფლების გადაუდებელი ამოცანაა.

References

1. Abuselidze, G. (2005). The prospects of modern budget revenue in the aspect of the new Tax Code. Journal of Social Economy, 1, 49-60.
2. Abuselidze, G. (2005). Peculiarities of formation and functioning of Georgian budget system in transient period. Author's abstract of dissertation, Tbilisi, 16.
3. Abuselidze, G. (2005). To optimize legal foundations of the budget system. Journal of Social Economy, 4, 72-80.
4. Abuselidze, G. (2006). Features of Formation and Functioning of Budgetary System of Georgia at the transitive stage. TB: Publishing House of Science.
5. Abuselidze, G. (2008). Taxation heaviness in Georgia and its optimization–for tax saving hints. Economics and Business, 4, 74-82.